\PassOptionsToPackage{dvipsnames}{xcolor}
\documentclass[10pt,conference]{IEEEtran}
\IEEEoverridecommandlockouts
\usepackage{cite}
\usepackage{amsmath,amssymb,amsfonts}
\usepackage{algorithmic}
\usepackage{graphicx}
\usepackage{textcomp}
\usepackage{xcolor}
\usepackage{balance}
\usepackage[all]{nowidow}
\usepackage{float}

\usepackage[hidelinks]{hyperref}
\usepackage[capitalise]{cleveref}

\usepackage{todonotes}
\definecolor{darkblue}{rgb}{0.0, 0.0, 0.55}
\begin{document}

\title{Towards Serverless Processing of Spatiotemporal Big Data Queries }

\author{\IEEEauthorblockN{Diana Baumann, Tim C. Rese, David Bermbach}
    \IEEEauthorblockA{\textit{Technische Universit\"at Berlin}\\
        \textit{Scalable Software Systems Research Group} \\
        \{diba,tr,db\}@3s.tu-berlin.de}
\thanks{© 2025 IEEE.  Personal use of this material is permitted.  Permission from IEEE must be obtained for all other uses, in any current or future media, including reprinting/republishing this material for advertising or promotional purposes, creating new collective works, for resale or redistribution to servers or lists, or reuse of any copyrighted component of this work in other works.}}


\maketitle

\IEEEpubidadjcol

\begin{abstract}
    Spatiotemporal data are being produced in continuously growing volumes by a variety of data sources and a variety of application fields rely on rapid analysis of such data.
    Existing systems such as PostGIS or MobilityDB usually build on relational database systems, thus, inheriting their scale-out characteristics.
	As a consequence, big spatiotemporal data scenarios still have limited support even though many query types can easily be parallelized. In this paper, we propose our vision of a native serverless data processing approach for spatiotemporal data:
	We break down queries into small subqueries which then leverage the near-instant scaling of Function-as-a-Service platforms to execute them in parallel.
	With this, we partially solve the scalability needs of big spatiotemporal data processing.
\end{abstract}

\begin{IEEEkeywords}
    Serverless Computing, Function-as-a-Service, Spatiotemporal Data, MapReduce
\end{IEEEkeywords}

\section{Introduction} 
\label{sec:introduction}
Nowadays, spatiotemporal data are being produced in exponentially growing volumes in a variety of application scenarios.
Examples of spatiotemporal data include mobility traces of cars, cyclists, pedestrians, planes, or ships in transport, movement patterns of wildlife in animal observation, Earth observation data, or satellite movement data~\cite{liang_survey_2024}.
Aside from data values such as a temperature measurement or a boolean ``something has a value here,'' spatiotemporal data contain additional space and time attributes~\cite{alam_survey_2022}.
Here, ``space attributes'' refers to locations usually indicated by latitude and longitude values and sometimes an additional geofence indicating a particular area for 2D spatial data or its 3D variant with an additional elevation level or 3D shapes.
Time attributes, in contrast, provide a temporal context, usually based on timestamps but possibly also based on timespans.
Such spatiotemporal data are usually encoded as either trajectory or raster data and stored in respective database systems.
In this paper, we focus on trajectory data.

While there are different ways for processing these data, e.g., in batch mode, many application domains rely on interactive analysis of big spatiotemporal data with queries that filter and aggregate by space and time attributes at the same time.
Existing solutions for this build on PostGIS or MobilityDB and, thus, inherit the limited scalability of the underlying relational PostgreSQL~\cite{rese2025evaluatingimpactspatialfeatures}.
Recent work by Bakli~et~al.~\cite{bakli2025mobilityDB} here represents the state-of-the-art.
Through complex sharding mechanisms and careful query planning, they managed to scale MobilityDB to up to eight machines -- an impressive feat.

\begin{figure}
    \centering
    \includegraphics[width=\columnwidth]{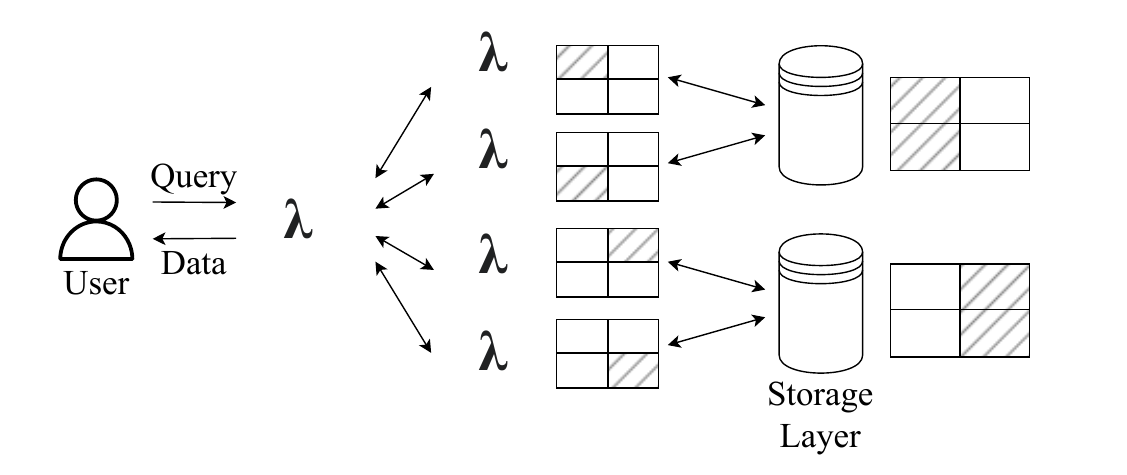}
    \caption{We leverage FaaS and shared storage for nodes to execute query in parallel on respective data subsets.}
    \label{fig:approach-sketch}
\end{figure}

At the same time, many query types can easily be broken down into vast numbers of subqueries operating on small data shards only, e.g., based on geographic regions or other spatial or temporal attributes, thus, parallelizing the query execution.
Today, the go-to solution for these ``embarrassingly parallel''~\cite{hellerstein2018serverless} tasks is Function as a Service (FaaS) in which developers write simple, stateless functions whose operation including on-demand scaling is then fully managed by the cloud provider~\cite{bermbach_future_2021}.

In this paper, we hence sketch out our vision of a serverless data processing approach for spatiotemporal big data analytics, thus, leverage the rapid on-demand scaling of FaaS function instances for parallelizing query processing.
We specifically do not make assumptions about how the data is structured, or whether it is being altered at some point during the analysis.
We address the challenges faced in data analytics with complex and read-heavy workloads where the sheer amount of data that needs to be analyzed is way beyond anything a single node can manage efficiently.
For this, we propose to break down queries into subqueries following the MapReduce model~\cite{dean_mapreduce_2008}, automatically deploy these on a FaaS platform, and merge the set of subresults into a single response (see~\cref{fig:approach-sketch}).
In our proposed approach, a dedicated coordinator function analyzes the query and then breaks it down based on spatial or temporal characteristics or a combination of both.

This idea is not completely new, in fact, our work was partially inspired by similar approaches for serverless processing of tabular data~\cite{lounissi2025funda,bodner2025skyrise}.
To our knowledge, it has not yet been proposed for spatiotemporal data -- a data type which needs to be dedicated sharding strategies which can consider spatial and temporal characteristics.
Existing sharding approaches are not directly applicable.
The reason for this is that storing spatial data already means reducing 2D or 3D data to a 1D sequence of values, which also makes indexing hard~\cite{rese2025evaluatingimpactspatialfeatures}, while the time context adds at least one more dimension.
Processing spatiotemporal big data on a single node, however, leads to slow response times as it stores and processes all data alone leading to nodes becoming slow which impacts any other workloads.

\section{Approach}
\label{sec:approach}
\begin{figure}
    \centering
    \includegraphics[width=\columnwidth]{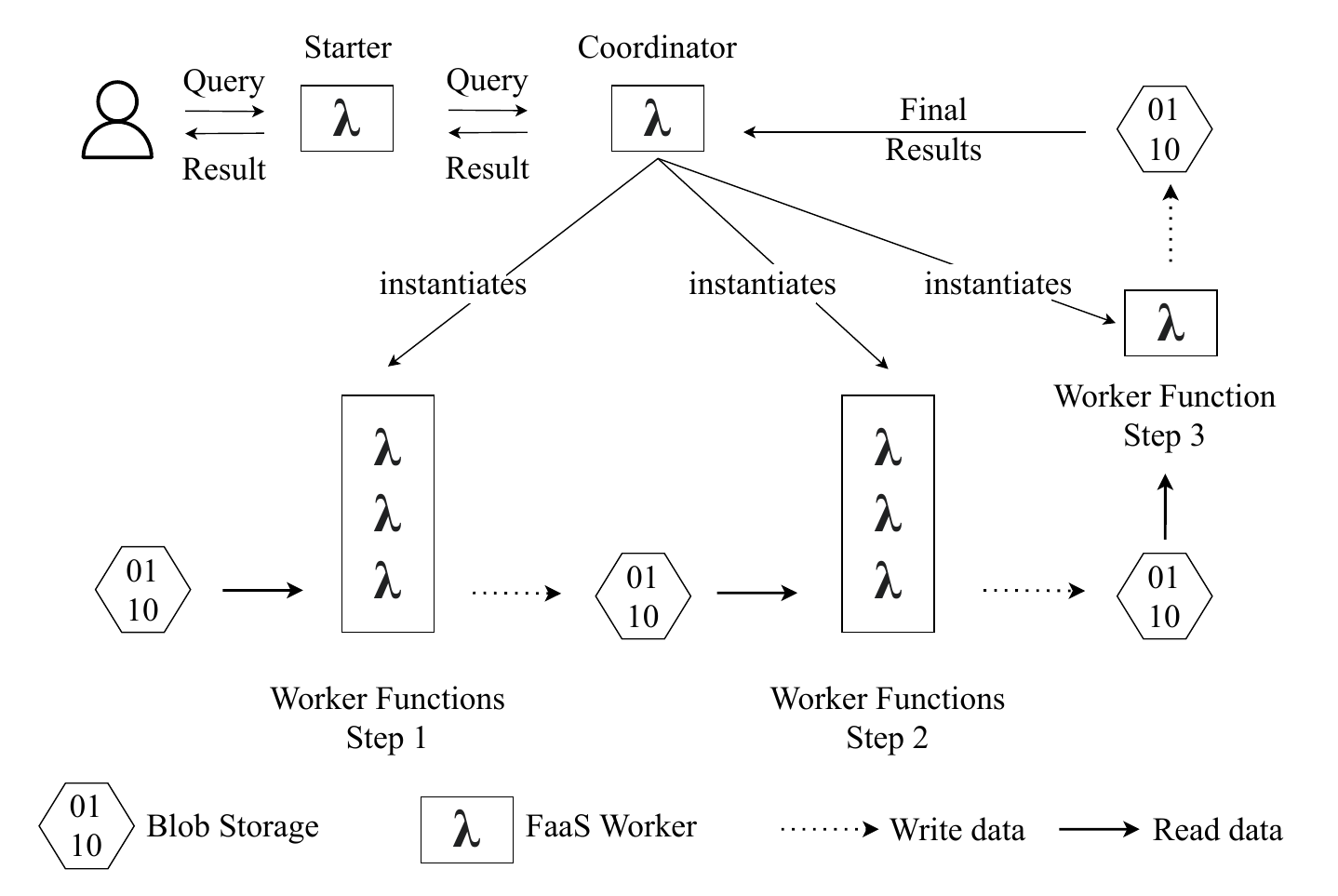}
    \caption{The coordinator manages waves of worker functions for each subquery with the amount of worker functions per wave dependent on the parallelizability of the subquery.}
    \label{fig:architecture-sketch}
\end{figure}

We propose to use a shared storage as data layer and shift query processing to FaaS applications.
Striving for a modular design which allows for later integration of our approach into diverse architectures, i.e., independent of the implemented execution framework, we use FaaS for processing and blob storage for storing results.
Our architecture design, depicted in \cref{fig:architecture-sketch}, is as follows.

A spatiotemporal query initiated by the client, e.g., ``Identify the flight-heaviest regions in the last six months” is our starting point.
A \texttt{starter} function provides an HTTP endpoint to receive the query and invokes the centralized \texttt{coordinator}.
The coordinator splits the query into subqueries and initializes the worker functions.
Queries such as \texttt{SELECT} can easily be distributed and executed on data shards, while others, e.g., \texttt{ORDER BY} or \texttt{AVG}, depend on context, i.e., they require the complete result of the previous subquery to return correct result and therefore can only be executed by single node.
This is why we design our architecture in a way that it supports all kinds of queries, parallelizable and non-parallelizable.
For that, the coordinator manages waves of worker functions for each subquery with the amount of worker functions per wave dependent on the parallelizability of the subquery.

The subqueries are executed on the assigned data shards.
Depending on the strategy chosen to shard the data, data shards can be, for example, flight data of a continent on Earth during the last two quarters.
This way, the data is sharded based on the spatial attribute \texttt{continents} and temporal attribute \texttt{quarter} simultaneously.
Further data sharding strategies may consider (i) the query type, e.g., spatial-only, temporal-only, or spatiotemporal, and (ii) processing load per node.

After the execution, the centralized orchestrator combines the results of each FaaS node to a unified answer to the query based on the complete dataset following the MapReduce principle.

\section{Conclusion \& Future Work}
\label{sec:conclusion}
Today, spatiotemporal data are being produced in continuously growing volumes and a variety of application fields rely on rapid analysis of such data.
    Existing systems such as PostGIS or MobilityDB, however, usually build on relational database systems, thus, inheriting their scale-out characteristics.
	As a consequence, \emph{big} spatiotemporal data scenarios still have limited support even though many query types can easily be parallelized.
		
	In this paper, we proposed our vision of a native serverless data processing approach for spatiotemporal data analysis with complex and read-heavy workloads:
	The proposed approach breaks down queries into small subqueries which then leverage the near-instant scaling of Function-as-a-Service platforms for parallel execution.

At the moment, we are working on implementing the proposed approach in a cloud-based prototype.
In future work, we are also planning to explore through experiments with spatiotemporal database systems and our approach how ``big'' data need to be to benefit from our approach:
We expect our approach to scale and offer near-constant latency which, however, is likely to be significantly higher than query latency in a single node deployment.

\balance

\bibliographystyle{IEEEtran}
\bibliography{bibliography}

\end{document}